\documentclass[preprint,aps,nofootinbib,preprintnumbers,amsmath,amssymb,11pt]{revtex4}
\usepackage{amsmath}
\usepackage{amssymb}
\usepackage{setspace}
\usepackage{amsfonts}
\usepackage{graphicx}% Include figure files
\usepackage{dcolumn}% Align table columns on decimal point
\usepackage{bm}% bold math
\usepackage{natbib}

\newcommand{\p}{\partial}
\newcommand{\half}{\frac{1}{2}}

\newcommand\nn{{\nonumber}}
\newcommand{\beq}{\begin{equation}}
\newcommand{\eq}{\end{equation}}

\def\bea{\begin{eqnarray}}
\def\ea{\end{eqnarray}}
\def\p{\partial}

\begin{document}

\title{Spatially Modulated Instabilities of Geometries with Hyperscaling Violation}

\author{Sera Cremonini$ ^{\,\clubsuit,\spadesuit}$}
\email{S.Cremonini@damtp.cam.ac.uk}
\affiliation{\it $ ^\clubsuit$
\it DAMTP, Centre for Mathematical Sciences, University of Cambridge, Wilberforce Road, Cambridge, CB3 0WA, UK \\
\it $ ^\spadesuit$ George and Cynthia Mitchell Institute for Fundamental Physics and Astronomy,
\it Texas A\&M University, College Station, TX 77843--4242, USA}
\author{Annamaria Sinkovics}
\email{sinkovics@general.elte.hu}
\affiliation{Institute of Theoretical Physics, MTA-ELTE Theoretical Physics Research Group,
E{\"o}tv{\"o}s Lor{\'a}nd University,  1117 Budapest, P{\'a}zm{\'a}ny s. 1/A,  Hungary}

\date{\today}

\begin{abstract}
We perform a study of possible instabilities of the infrared $AdS_2 \times \mathbb{R}^2$ region of solutions
to Einstein-Maxwell-dilaton systems which exhibit an intermediate regime of hyperscaling violation and Lifshitz scaling.
Focusing on solutions that are magnetically charged, we probe the response of the system to spatially modulated fluctuations, and identify
regions of parameter space in which the infrared $AdS_2$ geometry is unstable to perturbations.
The conditions for the existence of instabilities translate to restrictions on the structure of the gauge kinetic function and scalar potential.
In turn, these can lead to restrictions on the dynamical critical exponent $z$ and on the amount of hyperscaling violation $\theta$.
Our analysis thus provides further evidence for the notion that the true ground state of
`scaling' solutions with hyperscaling violation
may be spatially modulated phases.
\end{abstract}
\preprint{MIFPA-12-44}

\maketitle
\newpage
\tableofcontents
\newpage

%%%%%%%%%%%%%%%%%%%
\section{Introduction}
%%%%%%%%%%%%%%%%%%%

Within the framework of the gauge/gravity duality, geometries which give rise to interesting scaling behavior
continue to offer a rich testing ground for toy models of strongly correlated phenomena,
with potential applications to a number of condensed matter systems.
While spacetimes describing theories with a dynamical critical exponent $z$
have been studied for some time now (see \cite{Kachru:2008yh,Taylor:2008tg} for early realizations),
the notion of \emph{hyperscaling violation} has been explored holographically only recently.

Gravitational backgrounds which encode non-relativistic scaling
and non-trivial hyperscaling violation -- controlled by an exponent $\theta$ --
are supported by metrics of the form
\beq
\label{HVmetric}
ds_{d+2}^2 = r^{-\frac{2(d-\theta)}{d}} \left(-r^{-2(z-1)} dt^2 +dr^2 + d\vec{x}_d^2 \right) \, ,
\eq
which are not scale invariant but rather
transform as $ds \rightarrow \lambda^{\theta/d} ds$ under the scalings
$t \rightarrow \lambda^z t$, $x_i \rightarrow \lambda x_i$ and $r \rightarrow \lambda r$.
Solutions of this type have been seen to arise in simple Einstein-Maxwell-dilaton theories (see e.g.
\cite{Gubser:2009qt,Cadoni:2009xm,Charmousis:2010zz,Perlmutter:2010qu,Iizuka:2011hg,Gouteraux:2011ce,Huijse:2011ef})
thanks to a sufficiently non-trivial profile for the dilatonic scalar.

An appealing feature of the presence of a non-vanishing hyperscaling violating exponent $\theta$
is that it modifies the usual scaling of entropy with temperature, leading to $s\sim T^{(d-\theta)/z}$.
For this reason, geometries which realize $\theta = d- 1$ are of interest
for probing compressible states of matter (which may have `hidden' Fermi
surfaces \cite{Huijse:2011ef}), for which  $s \sim T^{1/z}$ independently of dimensionality.
In fact, solutions with $ \theta = d-1$ have been shown \cite{Ogawa:2011bz} to be associated with a
logarithmic violation of the area law of entanglement entropy,
\beq
\label{AlogA}
S_{ent} \sim A \log A \, ,
\eq
which is considered a signature of systems with a Fermi surface\footnote{Certain
current-current correlators \cite{Hartnoll:2012wm},
however, do not exhibit the finite momentum excitations expected in the presence of a Fermi surface, potentially undermining the interpretation
of these geometries as probing systems with a Fermi surface. This problem was circumvented in \cite{Hartnoll:2012wm} by suggesting that
the hyperscaling violating geometries
should be considered in an appropriate double scaling limit, in which both $\theta$ and $z$ approach infinity, with their ratio held fixed.}.
We refer the reader to e.g.
\cite{Dong:2012se,Hartnoll:2012wm,Ammon:2012je,Sadeghi:2012ix,Alishahiha:2012qu,Bueno:2012sd,Kulaxizi:2012gy,Adam:2012mw,Alishahiha:2012cm,Kim:2012pd,Edalati:2012tc,
Donos:2012yu,Donos:2012yi,Iizuka:2012pn,Gouteraux:2012yr,Gath:2012pg}
for various properties of these geometries, and attempts to classify the corresponding phases.

In the class of Einstein-Maxwell-dilaton theories which give rise to (\ref{HVmetric}), the scalar typically runs
logarithmically towards the horizon.
As a result, the `Lifshitz-like' hyperscaling violating solutions are believed to be a good description of the geometry
only in some intermediate near-horizon region, and are expected to be modified\footnote{Note however that there are cases
in which, after uplifting to higher dimensions, one recovers the expected `naive' scaling of thermodynamic quantities \cite{Gouteraux:2011ce}.
In such cases, the higher-dimensional embedding offers a potential resolution
of the singular behavior of the lower-dimensional zero temperature solutions.\\}
 in the deep infrared (IR).
The question of the possible IR completion of Lifshitz-like spacetimes
was examined in \cite{Harrison:2012vy} (see also
\cite{Horowitz:2011gh,Bao:2012yt} for related discussions in the context of pure Lifshitz systems without running couplings).
In the presence of hyperscaling violation, this issue was studied more recently in \cite{Bhattacharya:2012zu,Kundu:2012jn}
where -- for appropriate ranges of parameter space -- the solutions were shown to flow to $AdS_2 \times \mathbb{R}^2$ at the horizon, while approaching
$AdS_4$ in the ultraviolet (UV).
Thus, in the constructions of \cite{Bhattacharya:2012zu,Kundu:2012jn}, we see the emergence of an $AdS_2 \times \mathbb{R}^2$ description
in the deep IR, with its associated extensive zero temperature ground state entropy
in violation with the third law of thermodynamics\footnote{In \cite{Bhattacharya:2012zu} it was suggested that an IR $AdS_2$
description could be generated by including the types
of quantum corrections expected to become non-negligible as the dilaton drives the system towards strong coupling.
On the other hand, in \cite{Kundu:2012jn} it was
 the presence of both electric and magnetic fields which provided a stabilizing potential for the scalar field.
 Thus, while in both of these constructions the IR endpoint of the hyperscaling violating solutions is $AdS_2$, with the associated extensive
ground state entropy, the origin of the latter is of a different nature -- quantum mechanical in \cite{Bhattacharya:2012zu},
and classical in the dyonic system studied in \cite{Kundu:2012jn}.}.

In constructions of this type, however,
the near-horizon $AdS_2 \times \mathbb{R}^2$ geometry has been known to suffer from
\emph{spatially modulated} instabilities \cite{Donos:2011qt,Donos:2011pn}.
Thus, for the cases in which the (unstable) $AdS_2$ is the IR completion of an intermediate scaling region,
such instabilities appear to characterize the end-point of geometries which describe hyperscaling violation and anisotropic scaling.
As suggested in a number of places, this hints at the idea that the zero temperature ground state of these systems may in fact be
spatially modulated phases.
In fact, analogous (striped) instabilities have been studied very recently in \cite{Donos:2012yu},
in a particular $D=11$ SUGRA reduction which gave rise to purely magnetic hyperscaling violating solutions with $z=3/2$ and $\theta=-2$.
Moreover, analytical examples of striped phases were found recently in \cite{Iizuka:2012pn}
(see also \cite{Jokela:2012se,Rozali:2012es} for related work).

In this note, we would like to explore this idea further and -- motivated by \cite{Donos:2011qt,Donos:2011pn} --
examine the IR instabilities arising in Einstein-Maxwell-dilaton systems which allow for
intermediate scaling solutions with general values of $z$ and $\theta$.
In particular, we would like to identify conditions on the structure of the scalar potential and gauge kinetic function -- which
we initially take to be generic -- for which the geometry will be unstable to decay.
These conditions will then translate to restrictions on the value of the exponents $z$ and $\theta$ of the intermediate scaling regime -- as
well as on the remaining parameters of the theory.
As we will see, much like in \cite{Nakamura:2009tf,Donos:2011bh}, we will find a number of modulated instabilities at finite momentum,
lending evidence to the notion that $AdS_2$ should not describe the zero temperature ground state of the system -- rather,
the `scaling' solutions appear to be unstable to the formation of spatially modulated phases.
While our instability analysis is only a modest first step and is by no means general, we hope that it may offer some further insight
into the puzzle of the extensive ground state entropy
associated with the IR $AdS_2 \times \mathbb{R}^2$ completion of the `scaling' geometries, and of the true ground state of the theory.

The structure of the paper is as follows.
In Section \ref{Setup} we introduce our setup, focusing on properties of the background geometry.
Section \ref{SectionInstabilities} contains the linear perturbation and instability analysis.
We conclude in \ref{TheEnd} with a summary of results and a discussion of open questions.

%%%%%%%%%%%%%%%%%%%
\section{The Setup}
\label{Setup}
%%%%%%%%%%%%%%%%%%%

Our starting point is a four-dimensional Einstein-Maxwell-dilaton model of the form,
\beq
\label{instL}
\mathcal{L} = R - V(\phi) - 2 \left( \partial\phi\right)^2 - f(\phi) F_{\mu\nu} F^{\mu\nu} \, .
\eq
We are interested in potentials $V(\phi)$ and gauge kinetic functions $f(\phi)$ which allow
the geometry to be $AdS_2 \times \mathbb{R}^2$ in the deep infrared, and support an intermediate `scaling' region
with non-trivial $\{z,\theta\}$.
We choose the background gauge field to be that of a constant magnetic field,
\beq
\label{flux}
F = Q_m \, dx \wedge dy \; ,
\eq
and parametrize the metric,
which we take to be homogeneous and isotropic, by
\beq
\label{abcAnsatz}
ds^2 = L^2 \left( -a(r)^2 dt^2 + \frac{dr^2}{a(r)^2} + b(r)^2 d\vec{x}^2 \right) \, .
\eq
After simple manipulations, the equations of motion for the scalar and metric functions can be shown to reduce to
\bea
\label{aEOM}
\left(\p_r\phi\right)^2 &=&  -\frac{\p_r^2 b}{b} \, , \\
\label{cEOM}
4b^2 L^2 V(\phi) &=& -2 \, \p_r^2(a^2 b^2)  \, , \\
\label{bEOM}
4 f(\phi) \, Q_m^2 - 2 b^4 L^4 V(\phi) &=& 2 \, L^2 b^2 \p_r \left(b^2 \p_r(a^{2}) \right) \, , \\
\label{phiEOM}
4 f^\prime(\phi) \,  Q_m^2 + 2 \, b^4 L^4 \, V^\prime(\phi) &=& 8 \, L^2 b^2 \p_r \left(a^2 b^2 \, \p_r \phi \right) \, ,
\ea
where primes denote $^\prime \equiv \partial_\phi$.
Note that we have already made use of our flux ansatz.

%%%%%%%%%%%
\subsection{Conditions for the existence of $AdS_2 \times \mathbb{R}^2$ in the IR}
%%%%%%%%%%%

In order for the solutions to \eqref{instL} to reduce to $AdS_2 \times \mathbb{R}^2$ at the horizon,
the potential and the gauge kinetic function must satisfy appropriate conditions.
In particular, requiring the metric in the deed infrared to become of the form
\beq
\label{ads2metric}
ds^2 = L^2 \left( - r^2 dt^2 + \frac{dr^2}{r^2} + b^2 (dx^2+dy^2) \right) \, ,
\eq
with $b$ a constant,
and the scalar to also settle to a constant $\phi = \phi_h$ at the horizon, we find
\bea
\label{eq1}
&& 1+ \frac{L^2}{2} V(\phi_h) = \frac{f(\phi_h) Q_m^2}{b^4 L^2} \, , \\
\label{eq2}
&& 1+ L^2 V(\phi_h) = 0 \, , \\
\label{eq3}
&& \frac{2 f^\prime(\phi_h) Q_m^2 }{b^4 L^4} + V^\prime (\phi_h) = 0 \, .
\ea
From the last equation we learn that $V^\prime(\phi_h)/f^\prime(\phi_h) < 0$, after
imposing reality for the magnetic charge.
We can rearrange (\ref{eq1})--(\ref{eq3}) in a number of ways, and at this point it turns out to be
convenient to express them as:
\bea
\label{eq4}
V(\phi_h) &=& - \frac{1}{L^2} \, , \\
\label{eq5}
\frac{Q_m^2}{b^4 L^4} &=& \frac{1}{2 L^2 f(\phi_h)} \, , \\
\label{eq6}
\frac{f^{\, \prime}(\phi_h)}{f(\phi_h)} &=& \frac{V^{\, \prime}(\phi_h)}{V(\phi_h)} \, .
\ea
We will use these conditions throughout the instability analysis, to simplify background terms.

\subsection{Intermediate scaling regime}
%%%%%%%%%%%

Thus far we have kept the scalar potential and gauge kinetic function arbitrary, only subject to the requirement
that they should allow for $AdS_2 \times \mathbb{R}^2$ in the deep infrared. However, we are interested in solutions which flow to a geometry
characterized by non-trivial values of $z$ and $\theta$, in some intermediate portion of the spacetime.
Intermediate `scaling' solutions of this type can be engineered by choosing appropriately $V(\phi)$ and $f(\phi)$, and in particular
by taking them to be single exponentials, each characterized by its own exponent.
Thus, to guarantee the presence of a region exhibiting both anisotropic Lifshitz scaling and
hyperscaling violation,
\beq
\label{HVmetric2}
ds^2 = r^{-\frac{2(d-\theta)}{d}} \left(-r^{-2(z-1)} dt^2 +dr^2 + d\vec{x}_d^2 \right) \, ,
\eq
we will be interested in particular in the choice
\beq
\label{ranoutofnames}
f(\phi) = e^{2\alpha\phi} \, , \quad \quad V(\phi) = - V_0 e^{-\eta\phi} + \mathcal{V}(\phi) \, ,
\eq
where the first potential term is of the standard form needed to generate $\theta\neq 0$, and
$ \mathcal{V}(\phi) $ is assumed to be negligible in the intermediate scaling region.
The exponents $z$ and $\theta$ are then determined from the lagrangian parameters $\alpha$ and $\eta$ through the standard relations
(see e.g. \cite{Bhattacharya:2012zu} for magnetically charged solutions),
\beq
\label{thetazrelations}
\theta = -\frac{4\eta}{2\alpha-\eta} \, , \quad \quad
z=\frac{16+4\alpha^2-4\alpha\eta - 3\eta^2}{(2\alpha+\eta)(2\alpha-\eta)} \, .
\eq
Although our instability analysis will be carried out for a generic $V(\phi)$ and $f(\phi)$, only subject to the
infrared $AdS_2 \times \mathbb{R}^2$ requirements (\ref{eq1})--(\ref{eq3}), when connecting with the notion of hyperscaling violation we'll
adopt an ansatz of the form (\ref{ranoutofnames}).

%%%%%%%%%%%
\subsection{Explicit realizations}
%%%%%%%%%%%

Many of the explicit realizations in the literature of the interpolating geometries we have been discussing
are supported by a racetrack-type potential of the form
\beq
\label{racetrack}
V(\phi) =  -V_0 \, e^{-\eta\phi} + V_1 \, e^{\gamma \phi} \, ,
\eq
in terms of which the instability analysis of Section \ref{SectionInstabilities} will be particularly tractable.
While for now the constant $\gamma$ is left completely arbitrary, it will have to be such that
-- in \emph{some} part of the geometry -- the second exponential is subdominant.
In that region, then, the resulting hyperscaling violating, Lifshitz-like solution will be dictated entirely by $\alpha$ and $\eta$ through the
relations (\ref{thetazrelations}).
Here we touch on a few of the constructions in which (\ref{racetrack}) arises naturally, and
supports interpolating geometries with interesting scaling properties:
\begin{itemize}
\item
As an example, we would like to point out that
potentials of the type (\ref{racetrack}) arise in the \emph{equal scalars} case of the
$U(1)^4$ truncation \cite{Cvetic:1999xp}  of $D=4$ $SO(8)$ gauged supergravity studied in \cite{Donos:2011pn}.
This construction is particularly interesting as it gives rise to
magnetically charged solutions which flow from $AdS_2 \times \mathbb{R}^2$ near the horizon to $AdS_4$ at the boundary \cite{Donos:2011pn}.
%%%
%Moreover, as suggested in \cite{Harrison:2012vy,Bhattacharya:2012zu} and emphasized recently in \cite{Donos:2012yi},
%a regime of hyperscaling violation and Lifshitz-like scaling is expected to arise in these theories,
%as the solutions flow from the IR to the UV.
After setting the scalars all equal to each other, and taking three of the four gauge fields to be the same,
$F^{(2)} =F^{(3)}=F^{(4)}$,
the Lagrangian of \cite{Donos:2011pn} becomes (in our notation)
\beq
\label{rescaledLag}
\mathcal{L} = \half \left[ R -  2 \left(\p \phi\right)^2
-   e^{2\sqrt{3}\, \phi} F_{\mu\nu}F^{\mu\nu} -   \, e^{-\frac{2}{\sqrt{3}} \, \phi} \mathcal{F}_{\mu\nu}\mathcal{F}^{\mu\nu}
 + 6\left( e^{\frac{2}{\sqrt{3}} \, \phi} + e^{-\frac{2}{\sqrt{3}}\, \phi} \right)   \right] \, .
\eq
At the level of the background this action is of the form of (\ref{instL}), with
the gauge field kinetic term $ e^{-\frac{2}{\sqrt{3}} \, \phi} \mathcal{F}^2$ contributing to the (effective) scalar potential.
In this case the latter is of the racetrack form
$V =-V_0 e^{\frac{2}{\sqrt{3}} \, \phi}+ V_1 e^{-\frac{2}{\sqrt{3}} \, \phi}$.
Notice that if there is a region in the geometry in which the two conditions
\beq
\label{conditions}
1 \ll e^{\frac{4}{\sqrt{3}}\, \phi} \quad \text{and} \quad
e^{-\frac{2}{\sqrt{3}}\, \phi} \mathcal{F}^2 \ll e^{2\sqrt{3}\, \phi} F^2
\eq
are satisfied, the action would then reduce to the Einstein-Maxwell-dilaton system,
\beq
\mathcal{L} \approx  R -  2 \left(\p \phi\right)^2
-   e^{2\alpha\phi} F_{\mu\nu}F^{\mu\nu}  + V_0 e^{-\eta \phi}  \, ,
\eq
for the special values
\beq
\alpha = \sqrt{3} \, , \quad \eta = - \frac{2}{\sqrt{3}} \, ,
\eq
describing a `scaling' regime characterized by $z=3$ and $\theta=1$. 
Interesting scaling behavior was observed in systems of this type in
\cite{Donos:2012yi} where, however, the geometries were shown to be \emph{conformal} to $AdS_2$ in the infrared,
with interesting connections
to the double scaling limit of \cite{Hartnoll:2012wm}.
Finally, we note that our perturbation analysis of section \ref{SectionInstabilities}
applies to Einstein-Maxwell-dilaton theories with a single constant magnetic
field turned on, and therefore may not be directly applicable to the multi-charge systems studied in \cite{Donos:2011pn}.

\item
Racetrack potentials also arise in the (five-dimensional) Type IIB reduction studied in \cite{Kulaxizi:2012gy},
where a similar flow -- with an intermediate scaling regime -- was observed. There,
the near-horizon geometry was conformal to $AdS_2 \times R^3$. In the reduction of \cite{Kulaxizi:2012gy}, as well as in (\ref{rescaledLag}),
the parameters $\eta$ and $\gamma$
have the same sign and $V(\phi)$ acts as a trapping potential, as one may have naively expected.
Similar potentials have also been obtained by via dimensional reduction of e.g. Einstein-Maxwell theory \cite{Gouteraux:2011qh}.

\item
At the level of the background, the ansatz (\ref{racetrack}) also captures the dyonic setup of \cite{Kundu:2012jn},
with the scalar field  potential incorporating
the electric charge contribution to the flux term $f(\phi) F^2$.
In this case, then, we would read off that $\gamma = - 2\alpha$.
However, we emphasize that at the level of the perturbations our analysis will not directly apply to \cite{Kundu:2012jn},
since it is valid strictly for magnetically charged solutions.
For the dyonic case, one would have to take into account a more general set of perturbations (see e.g. the analysis of
\cite{Donos:2012yu}).
\end{itemize}

\noindent
{\bf Role of curvature}\\
We conclude this section by noting that
the curvature of the \emph{effective potential} for the scalar,
\beq
\label{curvature}
\frac{f^{\prime\prime}(\phi_h)}{f(\phi_h)} + V^{\prime\prime}(\phi_h) = 4\alpha^2 +
\left( - \eta^2 V_0 e^{-\eta\phi_h} + \gamma^2 V_1 e^{\gamma\phi_h} \right) \, ,
\eq
will play a key role in determining instabilities in Section \ref{SectionInstabilities}, as expected.
Since what we are after are constraints on the parameters
$\alpha,\eta$ and $\gamma$ entering the gauge kinetic function and scalar potential, we would like to eliminate the dependence of
$V^{\prime\prime}$ on $\phi_h$.
This can be easily done by using the requirement that the racetrack potential (\ref{racetrack}) allows for an $AdS_2 \times \mathbb{R}^2$ region
in the deep infrared.
In particular, making use of (\ref{eq4}) and (\ref{eq6}), we find
\beq
V_1 e^{\gamma\phi_h} = V_0 \, e^{-\eta\phi_h} - \frac{1}{L^2} \, , \quad \quad \quad
V_0 \, e^{-\eta\phi_h} = \frac{1}{L^2} \frac{\gamma-2\alpha}{\gamma + \eta} \, ,
\eq
which allow us to express $V^{\prime\prime}(\phi_h)$ in the more convenient form
\beq
\label{Vppracetrack}
V^{\prime\prime}(\phi_h) = \frac{1}{L^2} \Bigl(2\alpha\eta -\gamma(2\alpha+\eta) \Bigr) \, ,
\eq
controlled entirely by $\alpha,\eta$ and $\gamma$ as desired.
While we won't do it in full generality here, we note that this result can be expressed explicitly in terms of arbitrary $z$ and $\theta$ by
by inverting (\ref{thetazrelations}), which leads to\footnote{These relations were derived for
a solution that is magnetically charged.
For its electrically charged cousin, one must send $\alpha \rightarrow - \alpha$ in the expression (\ref{invertedrels}) for $\eta$.}
the following relations
\beq
\label{invertedrels}
\alpha^2 = \frac{(\theta-4)^2}{(\theta-2)(\theta-2z+2)} \, , \quad \quad \eta = \frac{2\theta\alpha}{\theta-4} \, .
\eq

\section{Spatially Modulated Instabilities\label{SectionInstabilities}}

Having introduced our setup, we are ready to examine the question of possible classical instabilities
of the IR $AdS_2\times \mathbb{R}^2$ region of the geometry.
In particular, we will study the response of the system to linear fluctuations and ask under what conditions,
if any, the $AdS_2$ BF bound is violated.
This will allow us to identify criteria for the existence of unstable modes, which will be dictated by the structure of the
gauge kinetic function and scalar potential.
In turn, these conditions will translate to \emph{restrictions on the values of $\{z,\theta\}$}
characterizing the hyperscaling violating `scaling' solutions which flow into $AdS_2\times \mathbb{R}^2$ in the infrared.

\subsection{Perturbation analysis}
In the deep IR, we take the background solution to the Einstein-Maxwell-dilaton system (\ref{instL}) to be described by a constant scalar
$\phi=\phi_h$
and an $AdS_2 \times \mathbb{R}^2$ metric parametrized by (\ref{ads2metric}).
Recall that we are interested in solutions supported by a constant background magnetic field $F_{xy}=Q_m$.
Following the discussion in \cite{Donos:2011qt}, we turn on the following set of
spatially modulated perturbations\footnote{Notice that we are working in radial
gauge, with the choice $\delta g_{ty}=0$, which is consistent for time-independent fluctuations. We thank Aristos Donos for clarifying this point.},
\bea
\delta g_{tt} &=& L^2 r^2 \, h_{tt}(r) \cos(k x) \, , \quad \delta g_{xx} = L^2 b^2 \, h_{xx}(r) \cos(k x) \, , \quad
\delta g_{yy} = L^2 b^2 \, h_{yy}(r) \cos(k x) \, , \nn \\
\delta A_y &=& a(r) \, \sin(k x) \, , \quad \quad \quad \delta \phi = w(r) \cos(k x) \, ,
\ea
anticipating that the instabilities will be driven by the finite momentum modulation.
Expanding the scalar and gauge fields equations of motion
\beq
\label{scalarEOM}
2c \; \Box \phi - V^{\prime} - f^{\, \prime} F_{\mu\nu}F^{\mu\nu} = 0 \, , \quad \quad \quad
\nabla_\mu \left(f(\phi) F^{\mu\nu} \right) = 0 \, , \nn
\eq
to linear order in the perturbations $\{ \delta\phi, \delta A_\mu \}$, we find
\bea
\label{gaugeeom}
&& \Box_{AdS_2} \, a = \frac{1}{L^2} \frac{k^2}{b^2} \, a + \frac{Q_m k}{2 b^2 L^2} \left( h_{tt}-h_{xx}-h_{yy} + \frac{2\bar{f}^\prime}{\bar{f}} w \right)\, , \\
\label{scalareom}
&& \Box_{AdS_2} \, w = \frac{1}{L^2}\left(\frac{k^2}{b^2} + \frac{\bar{f}^{\prime\prime}}{4\bar{f}} + \frac{\bar{V}^{\prime\prime}L^2}{4} \right) w
- \frac{1}{4L^2}\frac{\bar{f}^\prime}{\bar{f}} \left( h_{xx} + h_{yy}\right) -\frac{k \bar{f} Q_m \bar{V}^\prime }{L^2 b^4} \, a \, ,
\ea
where we are taking $\Box_{AdS_2} \equiv \frac{1}{L^2} \left(r^2 \partial_r^2 + 2 r \partial_r \right)$.
Barred objects denote background quantities, which are understood to be evaluated at $\phi=\phi_h$.
Expanding Einstein's equations to linear order in fluctuations we find
\bea
\label{einstein1}
&& \frac{k^2}{b^2} \, h_{tt} + h_{xx} + \left(1+ \frac{k^2}{b^2}\right) h_{yy}   -r \partial_r  \left(h_{xx} + h_{yy} \right)
 - \frac{4\bar{f} Q_m k}{L^2 b^4} \, a = 0 \, , \\
\label{einstein2}
&& k \, h_{tt} + k \, r \partial_r \left(h_{tt} + h_{yy} \right) - \frac{4\bar{f} Q_m}{L^2 b^2} \, r \partial_r a = 0 \, , \\
\label{einstein3}
&& \left( \Box_{AdS_2} - \frac{r\partial_r}{L^2} \right) \left(h_{xx} + h_{yy} \right) - \frac{1}{L^2} h_{xx}
-\frac{1}{L^2} \left( 1 + \frac{k^2}{ b^2} \right) h_{yy}
+ \frac{4\bar{f} Q_m k}{L^4 b^4} \, a = 0 \, , \\
\label{einstein4}
&& \Box_{AdS_2} \left( h_{yy}-h_{xx} \right) + \frac{k^2}{L^2 b^2}\, h_{tt} = 0 \, , \\
\label{einstein5}
&& \Box_{AdS_2} \left( h_{xx}+h_{yy}\right) + \left(2 \, \Box_{AdS_2} + \frac{2 r}{L^2} \partial_r  -\frac{k^2}{L^2 b^2} \right) h_{tt} +
\frac{2}{L^2} \left( h_{xx}+h_{yy}\right)
+ 4 \bar{V}^\prime w - \frac{8 \bar{f} Q_m k}{L^4 b^4} a =0 \, . \nn \\
\ea
For nonzero momentum\footnote{Notice that the $k=0$ case needs to be analyzed separately. We will return to this point shortly.},
this system can be reduced further by noticing that  (\ref{einstein1}) -- the $rr$ component of Einstein's equations --
is an algebraic equation for $h_{tt}$. Substituting  (\ref{einstein1}) into (\ref{gaugeeom}) and (\ref{einstein4}),
and noting that (\ref{scalareom}) and (\ref{einstein3}) do not involve $h_{tt}$,
we find the following system of equations for the remaining four perturbations,
\bea
\label{final1}
\Box_{AdS_2} h_{yy} &=& \frac{1}{L^2}\left(1+\frac{k^2}{b^2}\right) h_{yy} + \frac{1}{L^2} h_{xx} - \frac{4 k f Q_m }{L^4 b^4} \, a \, , \\
\label{final2}
\Box_{AdS_2} h_{xx} &=& \frac{r}{L^2} \left( h_{xx}^\prime + h_{yy}^\prime \right) \, , \\
\label{final3}
\Box_{AdS_2} a &=& \frac{1}{L^2}\left(1+\frac{k^2}{b^2}\right) a + \frac{k f^\prime Q_m }{L^2 b^2 f} \, w
- \frac{Q_m}{2L^2 k} \left[\left(1+\frac{k^2}{b^2}\right) h_{xx}+ \left(1+\frac{2k^2}{b^2}\right) h_{yy}\right] \nn \\
 && + \half \frac{Q_m}{L^2 k} r \p_r (h_{xx}+h_{yy})  \, , \\
\label{final4}
\Box_{AdS_2} w &=& \frac{1}{L^2}\left(\frac{k^2}{b^2} + \frac{f^{\prime\prime}}{4f} + \frac{V^{\prime\prime}L^2}{4} \right) w
- \frac{1}{4L^2}\frac{f^\prime}{f} \left( h_{xx} + h_{yy}\right) -\frac{k f Q_m V^\prime }{L^2 b^4} \, a \, ,
\ea
where we have dropped the barred notation for simplicity.
Finally, it is straightforward to check -- making use of (\ref{einstein2}) and (\ref{final1})-(\ref{final4}) --
that the $h_{tt}$ equation of motion (\ref{einstein5}) is satisfied.
At this stage it is clear from the structure of (\ref{final1})-(\ref{final4}) that these perturbations do not behave as scalars on $AdS_2$.
However, we expect that a more general \emph{time-dependent} ansatz will lead to a `proper' two-dimensional reduction,
as seen for example in the electrically charged cases studied in \cite{Donos:2011bh}
in which -- albeit in a slightly different context -- the perturbations fell into the nice form
$\Box_{AdS_2} \vec{v} = \mathcal{M}^2 \vec{v}$, with $\vec{v}$ denoting the vector of perturbations, and $\mathcal{M}^2$ the mass-squared matrix.

We now return to the zero momentum case, in which the perturbation equations (\ref{einstein1})-(\ref{einstein5}) reduce significantly.
In particular, after introducing
\beq
h_+ = h_{xx}+h_{yy}  \quad \quad \quad \text{and} \quad \quad \quad h_- = h_{xx}-h_{yy} \, ,
\eq
it is easy to see that $a(r)$ and $h_-(r)$ decouple from the remaining three perturbations,
leaving us with a simpler system for $\{h_+, h_{tt}, w\}$,
\bea
\label{a1}
 && \left(\Box_{AdS_2} - \frac{2}{L^2} \right) h_+ = 0 \, , \\
\label{a2}
 && \left(\Box_{AdS_2} + \frac{r\p_r}{L^2} \right) h_{tt} + \frac{2}{L^2}h_+ + 2 V^\prime w = 0 \, , \\
\label{a3}
 && \Box_{AdS_2}w - \frac{1}{4 L^2}\left( \frac{f^{\prime\prime}}{f} + L^2 V^{\prime\prime} \right) w + \frac{1}{4 L^2} \frac{f^\prime}{f}h_+  = 0 \, .
\ea

\subsection{Instabilities}

To approach the question of instabilities, we can now examine the spectrum of the scaling dimensions $\Delta$ associated with the
 perturbations, and ask whether they become complex in any regions of phase space,
signaling a violation of the $AdS_2$ BF bound.
In analogy with e.g. \cite{Donos:2011pn}, we expect that any potential instability will appear only at finite momentum $k$
(see \cite{Almuhairi:2011ws} for a study at zero-momentum).
Furthermore, since our goal here is to make a connection with the intermediate `scaling' part of the geometry,
we are particularly interested in how the gauge kinetic function $f$ and scalar potential $V$ affect the structure of
the various instabilities.\\

\noindent
\underline{{\bf Zero momentum case}}\\
Let us discuss briefly the zero momentum case, which corresponds to no spatial modulation.
It turns out to be convenient to package the perturbations $\{h_+, h_{tt}, w\}$ in a vector $\vec{v}$, so that
the system of equations (\ref{a1})-(\ref{a3}) can be put in matrix form, $\text{M} \vec{v} = 0$.
Assuming that the perturbations scale as $\vec{v} = \vec{v}_0 \, r^{-\delta}$,
with $\vec{v}_0$ a constant vector, the matrix takes the form
\beq
M = \begin{bmatrix}
   {\delta^2 - \delta -2} & 0 & 0  \\
  {2} & {\delta^2-2\delta} & {2 V^\prime L^2} &  \\
   {\frac{1}{4} \frac{f^\prime}{f} } & 0 & {\delta^2-\delta- \frac{1}{4} \left( \frac{f^{\prime\prime}}{f} + V^{\prime\prime}L^2 \right) }
    \end{bmatrix} \, .
\eq
The presence of classical instabilities will then be signaled by the roots of $\det(M)=0$ becoming complex.
In this case the determinant equation is easy to solve, and we find that the only solutions which are not
manifestly real\footnote{The remaining solutions are $\delta = 0, -1, 2$.}
 are of the form
\beq
\delta_\pm = \half \pm \half \sqrt{1+L^2 V^{\prime\prime} + \frac{f^{\prime\prime}}{f}} \, .
\eq
However, notice that $L^2 V^{\prime\prime} + \frac{f^{\prime\prime}}{f}$ is nothing but the curvature of the effective potential
$V_{eff}$. Thus, for $V_{eff}^{\prime\prime}> -\frac{1}{L^2} $ we don't see any instabilities at zero momentum. In particular,
if
the dilatonic scalar sits at a minimum of the effective potential,
as it does in the $AdS_2 \times \mathbb{R}^2$ background solution,
we have $V_{eff}^{\prime\prime}>0$. \\

\noindent
\underline{{\bf Finite momentum}}\\
We now move on to the spatial modulation case, for which the momentum is no longer vanishing.
As before, the perturbations $\{h_{xx},h_{yy}, a, w\}$ can be packaged in a vector $\vec{v} = \vec{v}_0 \, r^{-\delta}$,
in terms of which  the system of equations (\ref{final1})-(\ref{final4}) takes the form $\text{M} \vec{v} = 0$,
with
\beq
M = \begin{bmatrix}
  {-1}   &  {\delta^2 - \delta -(1+\frac{k^2}{b^2})} & \frac{4fQ_m k}{L^2 b^4} & 0  \\
   {\delta^2} & {\delta} & 0 & 0 \\
   {\frac{Q_m}{2k} \left(1+\frac{k^2}{b^2} +\delta  \right)} & {{\frac{Q_m}{2k} \left(1+2\frac{k^2}{b^2} +\delta  \right)}} &
    {\delta^2-\delta-(1+\frac{k^2}{b^2})} & -\frac{f^\prime Q_m k }{b^2 f} \\
    {\frac{f^\prime}{4f}} & {\frac{f^\prime}{4f}} & \frac{f Q_m V^\prime}{b^4}k &
    {\delta^2-\delta- (\frac{k^2}{b^2} + \frac{f^{\prime\prime}}{4f} + \frac{V^{\prime\prime}L^2}{4} )}
    \end{bmatrix} \, .
\eq
For arbitrary momentum $k$, the roots of $\det(M)=0$ are significantly more complicated.
For the sake of simplicity, we will therefore approximate the determinant by expanding it for small $k$,
neglecting terms of ${\cal O} (k^4)$ and higher.
The large $k$ limit will work in an analogous manner, although we expect that it will lead to different bounds on the
parameters of the system. Since here we are not after the most general set of instabilities,
we will content ourselves with a small $k$ approximation.
To this order in momentum, the solutions for $\delta$ which are not manifestly real take the form
\beq
\delta_{1,2,3,4} = \half \pm \sqrt{P_1 \pm \sqrt{P_2}} \; ,
\eq
with
\beq
P_1 \equiv \frac{5}{4} + \frac{1}{8}  \left( \frac{f^{\prime\prime}}{f} + L^2 V^{\prime\prime}\right) + \frac{3 k^2}{2b^2} \, ,
\eq
and
\beq
\label{P2exp}
P_2 = \left[1 - \frac{1}{8}\left(L^2 V^{\prime\prime}+\frac{f^{\prime\prime}}{f}\right) \right]^2
+  \left[1  -\frac{1}{8}
\left(  L^2 V^{\prime\prime} + \frac{f^{\prime\prime}}{f} \right) + \half\left(\frac{f^\prime}{f}\right)^2 \right] \frac{k^2}{b^2}
+ {\cal O}(k^4) \, .
\eq
At this point, the presence of an instability can be determined by asking whether the entire quantity
$P_1 \pm \sqrt{P_2}$ becomes negative, as this corresponds to a complex scaling dimension.
This will then lead to conditions on the structure of the gauge kinetic function and scalar potential.
In the small $k$ limit, we have
\bea
P_1 \pm \sqrt{P_2} &=&  \frac{5}{4} +\frac{1}{8} \, V_{eff}^{\prime\prime} \pm \left( 1- \frac{1}{8} V_{eff}^{\prime\prime}\right) +
\frac{k^2}{b^2} \left[ \frac{3}{2} \pm \half \pm  \frac{2}{8 - V_{eff}^{\prime\prime}} \frac{f^{\prime\; 2}}{f^2}  \right] \, ,
\ea
where we have made use of the more compact expression
\beq
V_{eff}^{\prime\prime} = \frac{f^{\prime\prime}}{f} + L^2 V^{\prime\prime} \, .
\eq
It is now easy to see that the root
\beq
P_1 + \sqrt{P_2} = \frac{9}{4} +  \frac{2 k^2}{b^2} \left[1 + \frac{1}{8 - V_{eff}^ {\prime \prime }} \left(\frac{f^\prime}{f}\right)^2  \right]
\eq
will become negative -- signaling an instability -- when
\beq
\label{GenCond1}
8 < V_{eff}^ {\prime \prime } < 8 + \left(\frac{f^\prime}{f}\right)^2 \, ,
\eq
for the range of $k$ for which the contribution from the ${\cal O}(k^2)$ term
dominates over the leading zero momentum term.
Similarly, by inspecting the other root,
\beq
\label{P1mP2}
P_1 - \sqrt{P_2} = \frac{1}{4} \left(1+V_{eff}^ {\prime \prime }\right) + \,
\frac{k^2}{b^2} \left[1 - \frac{2}{8-V_{eff}^ {\prime \prime }}  \left(\frac{f^\prime}{f}\right)^2  \right]
\eq
we see that the $k$-dependent term will be negative
when the curvature of the effective potential is in the range
\beq
\label{GenCond2}
8 -2 \left(\frac{f^\prime}{f}\right)^2 < V_{eff}^ {\prime \prime } < 8   \, .
\eq
Notice that, as we discussed earlier, the zero-momentum contribution to (\ref{P1mP2}) is positive when
$V_{eff}^ {\prime \prime }>-1$, and in particular when
the $AdS_2 \times \mathbb{R}^2$ background solution sits at a minimum of the effective potential.
Combining the two expression (\ref{GenCond1}) and (\ref{GenCond2}),
we find that spatially modulated instabilities will be present -- in an appropriate momentum range -- when the curvature of the effective potential
$V_{eff}^ {\prime \prime }= \frac{f^{\prime\prime}}{f} + L^2 V^{\prime\prime}$,
evaluated at $\phi=\phi_h$, is in the window
\beq
\label{GenInst}
8 -2 \left(\frac{f^\prime}{f}\right)^2 <
\frac{f^{\prime\prime}}{f} + L^2 V^{\prime\prime}
< 8 +  \left(\frac{f^\prime}{f}\right)^2 \, .
\eq
We emphasize that this relation is valid for a \emph{generic} scalar potential and gauge kinetic function, only subject to the requirement
that they lead to an $AdS_2 \times \mathbb{R}^2$ region in the far infrared.

However, to connect this discussion with the intermediate hyperscaling violating regime,
we will now assume that they are of the form of (\ref{ranoutofnames}), which we recall here for convenience,
$$f(\phi) = e^{2\alpha\phi} \, , \quad \quad \text{and} \quad \quad V(\phi) = - V_0 e^{-\eta\phi} + \mathcal{V}(\phi) \, .$$
In terms of these,  the finite-$k$ instability condition (\ref{GenInst}) can be expressed, making use of (\ref{eq4}), as
\beq
\label{GenInst2}
8 - 12 \alpha^2 < -\eta^2 + L^2 \left( \mathcal{V}^ {\prime \prime }(\phi_h) - \eta^2 \mathcal{V}(\phi_h) \right) < 8 \, .
\eq
In particular, for the simple racetrack potential (\ref{racetrack}), it reduces to
\beq
\label{RacetrackCond}
8 - 12 \alpha^2 < 2\alpha(\eta-\gamma) -\eta\gamma < 8 \, ,
\eq
where we made use of (\ref{Vppracetrack}).
Finally, one can reexpress this condition entirely in terms of $z$, $\theta$ and $\gamma$ by recalling that
$\eta = \frac{2\theta\alpha}{\theta-4}$ and $\alpha^2 = \frac{(\theta-4)^2}{(\theta-2)(\theta-2z+2)}$.
Since this is rather cumbersome -- but straightforward to obtain -- we won't include it here.\\

We conclude this discussion by examining a few special cases, including some of the constructions we
discussed in Section \ref{Setup}:
\begin{itemize}
\item
To describe $\theta=1$, the case associated with log violations of the entanglement entropy area law,
we need to set $\eta = -2\alpha/3$ and $\alpha^2 = 9/(2z-3)$. Notice that reality of $\alpha^2$ tells us that
$z>3/2$.  With these values, we find that we have instabilities whenever the racetrack potential parameter $\gamma$ lies in the range
\beq
\frac{3-4z}{\sqrt{2z-3}} < \gamma < \frac{2(15-2z)}{\sqrt{2z-3}} \, .
\eq
In particular, the special case in which $\gamma=\eta$ (as e.g. in a $\cosh\phi$ potential)
will be unstable as long as $z$ is in the range $ \frac{5}{4} < z < 8$.
\item
When $\eta = \gamma$, which is needed to support a $\cosh\phi$ potential, (\ref{RacetrackCond}) simplifies to\footnote{One
side of the inequality is satisfied trivially.}
\beq
\eta^2 - 12 \alpha^2 + 8 < 0 \, .
\eq
Written in terms of $z$ and $\theta$, the condition for instabilities becomes
\beq
\frac{6 \, \theta -14 - z(\theta-2)}{(\theta-2)(\theta-2z +2)} < 0 \, .
\eq
For the case with $\theta=1$, one then find that $z<8$ will correspond to the presence of unstable modes.
\end{itemize}
While here we have examined only a few explicit examples, the same logic can be applied to
systems described by (\ref{instL}), under the assumption of a constant background magnetic field, by making use of the general condition (\ref{GenInst}).

\section{Discussion}
\label{TheEnd}

One of the appealing features of `scaling' geometries with non-trivial $\{z,\theta\}$
is that they have zero entropy at zero temperature, in agreement with the third law of thermodynamics.
On the other hand, $AdS_2 \times \mathbb{R}^2$ is known to suffer from an extensive zero temperature ground state entropy,
an indication that the theory may be unstable.
This raises the natural question of what is the ultimate IR fate of the $\{z,\theta\}$ scaling solutions whose infrared completion is
$AdS_2 \times \mathbb{R}^2$ -- and in particular, of what is the true zero temperature ground state of the field theoretical systems they describe.
With these motivations in mind, in this note we have studied a class of instabilities of
magnetically charged $AdS_2 \times \mathbb{R}^2$ geometries arising in Einstein-Maxwell-dilaton theories
which can support an \emph{intermediate} regime of hyperscaling violation and Lifshitz scaling\footnote{The analysis of \cite{Iizuka:2013ag}
identifies striped instabilities by examining the scaling geometries directly (without assuming a 
flow to $AdS_2$ in the IR), and is therefore complementary to ours.}.

In particular, by examining the response of the system to spatially modulated fluctuations, we have identified
conditions for the existence of instabilities in the far infrared,
sensitive to the structure of the scalar potential $V(\phi)$ and gauge kinetic function $f(\phi)$ in the theory.
As in a number of examples in the literature, in this context the instabilities appear at finite $k$ and are therefore
intimately tied to the spatial modulation.
Working in a small $k$ approximation for simplicity, we have seen that -- in an appropriate momentum range --
the system is unstable to modulated perturbations when the conditions
\beq
\label{CondConclusions}
8 -2 \left(\frac{f^\prime}{f}\right)^2 <
\frac{f^{\prime\prime}}{f} + L^2 V^{\prime\prime}
< 8 +  \left(\frac{f^\prime}{f}\right)^2 \, ,
\eq
are satisfied in the far infrared, where the geometry is $AdS_2 \times \mathbb{R}^2$.

For models which can give rise to intermediate `scaling' solutions,
the conditions (\ref{CondConclusions}) can then be mapped to restrictions on $z$ and $\theta$, as well as on the remaining parameters in the theory.
A particularly tractable example is that of a racetrack potential
of the form $V(\phi) = - V_0 e^{-\eta\phi} + V_1 e^{\gamma\phi}$, for which instabilities arise
when
\beq
8 - 12 \alpha^2 < 2\alpha(\eta-\gamma) -\eta\gamma < 8 \, .
\eq
As a simple application of this relation, we note that
when $\gamma=\eta$, a choice which accommodates a $\cosh\phi$ potential
as well as many of the constructions in the literature, the system will be unstable when
\beq
\frac{6 \, \theta -14 - z(\theta-2)}{(\theta-2)(\theta-2z +2)} < 0 \, .
\eq
For the $\theta=1$ case, of particular interest as it is tied to log violations of the area law of entanglement entropy,
this condition translates to the restriction $z<8$.
The same procedure can then be applied to more non-trivial models by using (\ref{CondConclusions}).

We emphasize that our instability analysis is in no way exhaustive --
it applies to Einstein-Maxwell-dilaton theories with a background magnetic field only\footnote{For the case of a background electric field, 
see \cite{Cremonini:2013epa}.},
and does not involve the most general class of perturbations.
However, it provides further evidence for the notion that solutions with
hyperscaling violation are unstable to decay in the deep infrared -- and in particular, to the formation of spatially modulated phases --
and that $AdS_2 \times \mathbb{R}^2$ should not describe the true ground state of the theory at zero temperature.

A number of questions remain open.
First of all, we expect that a more general instability analysis
-- for arbitrary momentum and
time-dependent fluctuations -- may yield \emph{more stringent constraints} on $z$ and $\theta$.
It would be interesting to explore if, by turning on a background electric field and additional charges, the system might
exhibit a \emph{qualitatively different} behavior.
With a more general analysis in mind,
we wonder whether there is any notion of `universality' for the values of $\{z,\theta\}$
associated with an unstable infrared $AdS_2$ region.
Also, it is natural to ask how the structure of instabilities is modified for
`scaling' solutions which can be embedded in higher dimensions, and how this story ties into
some of the recent work on classifying hyperscaling violation,
and generating it from dimensional reduction \cite{Gouteraux:2011qh,Donos:2012yi,Donos:2012yu,Gouteraux:2012yr,Iizuka:2012pn}.
Interestingly, solutions with an intermediate scaling regime which approach a \emph{supersymmetric} $AdS_2 \times \mathbb{R}^2$
geometry in the IR have recently been found \cite{Donos:2012yi}, with the emergent infrared SUSY suggesting that they may in fact be stable.
Finally, there is the related question of what one could learn,
if anything, by applying a similar instability analysis to geometries which are \emph{conformal}
to $AdS_2 \times \mathbb{R}^2$ and support the double scaling limit of \cite{Hartnoll:2012wm}.
We leave these questions to future work.

%%%%%%%%%%%%%%%%%%%%%%%%%%%%%%%%%%%%%%%%%%%%%%%%%%%%%%%%%%%%%%%%%%%%%%%%%
\vskip0.5cm
\noindent
{\bf \large{Acknowledgement}}\\
%%%%%%%%%%%%%%%%%%%%%%%%%%%%%%%%%%%%%%%%%%%%%%%%%%%%%%%%%%%%%%%%%%%%%%%%%
We are particularly grateful to Aristos Donos, Claude Warnick and Scott Watson for many illuminating conversations,
and to Jyotirmoy Bhattacharya for initial collaboration.
We also thank Jerome Gauntlett, Sean Hartnoll, Liza Huijse, Shamit Kachru and Sandip Trivedi
for comments on the draft, and Yi Pang and Malcolm Perry for useful conversations.
The work of S.C. has been supported by the Cambridge-Mitchell Collaboration in Theoretical
Cosmology, and the Mitchell Family Foundation.

\end{document}